\documentstyle[11pt]{article}
\advance\textwidth by 3.5 truecm
\advance\textheight by 3 truecm
\def\koer#1{\hbox{\bf #1}}
\def\liealg#1{\hbox{\bf #1}}
\def\labelpr#1{\label{#1}}
\def\verw#1{(\ref{#1})}
\def\nn{\nonumber}
\def\bee{\begin{equation}}
\def\ene{\end{equation}}
\def\bea{\begin{eqnarray}}
\def\ena{\end{eqnarray}}
\def\bdis{\begin{displaymath}}
\def\edis{\end{displaymath}}
\def\agl{\, = \, }
\def\abgl{\quad = \quad }
\def\tgl{& = & }

\def\abdef{\quad := \quad }

\def\teqverw#1{& {\buildrel {\scriptstyle \verw{#1}}\over = } &}
\def\eqverw#1{\quad {\buildrel {\scriptstyle \verw{#1}}\over = } \quad}
\def\Sequ{\quad \buildrel \mbox{\small s} \over = \quad }
\def\gefordert{\; \buildrel ! \over = \;}
\newcommand{\qed}{\\ \rightline{$\Box$}\\ }
\def\p{\; +\; }
\def\m{\; -\; }
\def\quot#1#2{ \textstyle \, {{ #1}\over {#2}} \, }
\begin{document}
\begin{titlepage}
\vspace{50pt}
%
%
\title {\huge The exponential map of GL(N) \\[40pt]}
\author{{\Large Alexander Laufer}
       \thanks{{\bf e-mail} Alexander.Laufer@uni-konstanz.de}    \\[15pt]
                Department of physics   \\
                University of Konstanz  \\
                P.O. 5560 M 678         \\[10pt]
                78434 KONSTANZ          \\[30pt] }
\date{April 9, 1996}
%
%
%
\end{titlepage}
\vspace{70pt}
\maketitle
\begin{abstract}
A finite expansion of the exponential map for a
$N\times N$ matrix is presented.
The method uses the Cayley-Hamilton theorem for writing the higher  
matrix
powers in terms of the first N-1 ones. The resulting sums over the
corresponding coefficients are rational functions of the eigenvalues
of the matrix.
\end{abstract}
%
%
%
%
\newpage
\section{Introduction}
In the Lie theory of groups and their corresponding algebras the  
exponential
map is a crucial tool because it gives the connection between a Lie  
algebra
element $H\, \in $ \liealg{g} and the corresponding Lie group element
g $\in $ G
\bdis
\exp \quad : \quad \left.
\begin{array}{rcl}
                   \liealg{g} & \longrightarrow & \mbox{G}   \\
                         H    & \longmapsto     & \mbox{g}
\end{array}\right.
\edis
(For details see \cite{helga} and \cite{bzl1,bzl2} and ref.~therein).
In some low dimensional cases, like SU(2) and SO(3) the explicit  
expansion
of the exponential map is known. Some years ago the exponential map for
the Lorentz group was given by W. Rodrigues and J. Zeni  
\cite{zeni90,zeni92}.
For the higher dimensional groups $SU(2,2) $ and $O(2,4)$ a method  
for the
expansion was developed by A.O. Barut, J. Zeni, and A.L.  
\cite{bzl1,bzl2}.
The subject of the present paper is a generalization of the method
developed in \cite{bzl1, bzl2} to the general linear groups GL(N).
The result will be a method to calculate the exponential of a quadratic 
matrix $H$, where only rational functions of the eigenvalues of $H$ and
the first N-1 powers of $H$ are involved. The key points are the
Cayley-Hamilton theorem and the introduction of a multiplier $m$.
\\
The organization of this paper is as follows. First the method is shown 
in the low dimensional case SU(3) which is a group occurring quite often 
in physics.
Then the general case of the expansion of the exponential map for a
$N\times N $ matrix is presented. The method uses eignvalues

The only real problem that remains is the determination of the  
eigenvalues
of the matrix $H$. Throughout the paper
we assume that the groups GL(N) are represented as N-dimensional matrices
and that the eigenvalues of $H$ are all different
(cf. remark in appendix \ref{suNrem1}).
\\
Some possible applications of these results will be presented in a
future paper.
\section{The exponential map for the group SU(3) \labelpr{secexposu3}}
The group SU(3) is used in several branches of physics. The best known
application is the model of the strong interaction (see e.g.  
\cite{chengli}).
For this reason and because its a good exercise to follow the steps  
of the
general method we will show the exponential mapping of SU(3) in  
great detail.
The calculations depend in some points on the fact that we consider  
a special
group, i.e., the sum over the eigenvalues of the generator  
vanishes. There
is no conceptional problem to extend the method to U(3).
Like in the other cases (cf. \cite{bzl1,bzl2}) a typical element
U$\, \in $ SU(3) can be written as exponential of the generator
\bee
U \abgl e^H \abgl \sum_{n=0}^{\infty} \, \quot{1}{n!} \, H^n
\qquad \mbox{with} \qquad  H\, \in \,\liealg{su}(3) \, .
\labelpr{su3expo1}
\ene
The Cayley-Hamilton theorem and the iterated form in this case read
\bdis
H^3 \abgl b_0 \, H \p c_0 \qquad \mbox{and} \qquad
H^{3+i} \abgl a_i \, H^2 \p b_i \, H \p c_i
\edis
where the coefficients $b_0 $ and $c_0 $ are functions of the  
eigenvalues of
the eigenvalues $x , y, z$ of $H$. They satisfy the recurrence relations
\bee
a_{i+1} \agl b_{i} \; , \qquad b_{i+1} \agl a_{i} \, b_0 \p c_i \; ,
 \qquad c_{i+1} \agl a_{i} \, c_0 \, . \labelpr{recrelsu3}
\ene
Hence the coefficients $a_{i} $ satisfy
\bee
a_{i+1} \agl a_{i-1} \, b_{0}  \p   a_{i-2} \, c_{0} \labelpr{recrelsu3a}
\ene
with the first few values
\bdis
a_0 \agl 0, \qquad  a_1 \agl b_0 , \qquad a_2 \agl c_0 , \qquad
a_3 \agl {b_0}^2 , \qquad a_4 \agl b_0\, c_0 \; .
\edis
The explicit form of $b_0 $ and $c_0 $  can easily be
derived from the secular equation
\bdis
0 \agl \left( \lambda \m x \right) \, \left( \lambda \m y \right) \,
  \left( \lambda \m z \right) \agl \lambda^3 \m  (
\underbrace{x \p y\p z }_{a_0 \, =\, 0} ) \, \lambda^2  \p
( \underbrace{x\, y \p x\, z \p y\, z }_{-\, b_0} ) \, \lambda  \m
\underbrace{x \, y\, z }_{c_0} \, .
\edis
The leading coefficient $a_0$ vanishes since the generator $H$ is  
traceless,
i.e., $ x + y + z \agl 0 $. The second coefficient can also be written as
$b_0 \agl \quot{1}{2} \left( x^2 \p y^2 \p z^2 \right) $. There are  
also some
nice relations
\bdis
b_0 \, x \p c_0 \agl x^3  \; , \qquad  b_0 \, y \p c_0 \agl y^3  \; , 
\qquad  b_0 \, z \p c_0 \agl z^3   \; .
\edis
The idea now is to use the Cayley-Hamilton theorem for writing the sum
\verw{su3expo1} like
\bea
U \tgl I_4 \p H \p \quot{1}{2} H^2 \p \sum_{n=0}^{\infty}  
\quot{1}{(n+3)!}
    \, H^{n+3} \nn \\
\tgl I_4 \p H \p \quot{1}{2} H^2 \p \sum_{n=0}^{\infty}
 \quot{1}{(n+3)!} \left( a_n \, H^2 \p b_n \, H \p c_n \right)  \nn 
\ena
This form contains only sum over rational functions, there are no higher
powers of the generator present anymore. The next step is now to find an
analytic expression for the sums over the coefficients.
\\
A convenient form of the functions $a_n \, ,\, b_n $, and $ c_n $  
can be obtained if we introduce the multiplier
\bdis
m \agl \left( x \m y \right) \, \left( x \m z \right) \,
   \left( y \m z \right)
  \agl  \left( x^2 \,  \left( y \m z \right) \p y^2 \, \left( z \m  
x \right)
               \p z^2 \,  \left( x \m y \right) \right)   \; .
\edis
Then we get for the group element
\bdis
m\, U \agl  m \left( I_4 \p H \p \quot{1}{2} H^2 \right) \p
  \left[ \sum_{n=0}^{\infty} \frac{m\, a_n }{(n+3)!} \right] \, H^2 \p
   \left[ \sum_{n=0}^{\infty} \frac{m\, b_n }{(n+3)!} \right] \, H \p
  \left[ \sum_{n=0}^{\infty} \frac{m\, c_n }{(n+3)!} \right]  \; .
\edis
It can easily be shown that the following form for the coefficients 
satisfy the recurrence relations  \verw{recrelsu3} and
\verw{recrelsu3a}
\bea
m\, a_n \tgl \left( y \m z \right) \, x^{n+3} \p  \left( z \m x \right) 
    \, y^{n+3} \p  \left( x \m y \right) \, z^{n+3} \nn \\
m\, b_n \tgl \left( y \m z \right) \, x^{n+4} \p  \left( z \m x \right) 
    \, y^{n+4} \p  \left( x \m y \right) \, z^{n+4} \nn \\
m\, c_n \tgl y\, z \left( y \m z \right) \, x^{n+3} \p x\, z
  \left( z \m x \right) \, y^{n+3} \p  x\, y \left( x \m y \right)  
\, z^{n+3}
  \nn
\ena
The three sums are now
\bea
\left[ \sum_{n=0}^{\infty} \frac{m\, a_n }{(n+3)!} \right] \tgl
   \left( y \m z \right) \, e^{x} \p  \left( z \m x \right)
    \, e^{y} \p  \left( x \m y \right) \, e^{z} \m \quot{1}{2} m \nn \\ 
\left[ \sum_{n=0}^{\infty} \frac{m\, b_n }{(n+3)!} \right] \tgl
  x\, \left( y \m z \right) \, e^{x} \p y\, \left( z \m x \right)
    \, e^{y} \p z\, \left( x \m y \right) \, e^{z} \m  m  \nn \\
\left[ \sum_{n=0}^{\infty} \frac{m\, c_n }{(n+3)!} \right]  \tgl
  y\, z\, \left( y \m z \right) \, e^{x} \p x\, z\, \left( z \m x  
\right)
    \, e^{y} \p  x\, y\, \left( x \m y \right) \, e^{z}   \nn
\ena
Finally, we get the expansion of a SU(3) group element
\bee
\fbox{ \parbox{12cm}{
\bea
m\, U \tgl \left[  y\, z\, \left( y \m z \right) \, e^{x} \p x\, z\, 
   \left( z \m x \right)\, e^{y} \p  x\, y\, \left( x \m y \right)  
\, e^{z}
                \right] \; I_3 \nn \\
 && +\;  \left[  x\, \left( y \m z \right) \, e^{x} \p y\,
   \left( z \m x \right)\, e^{y} \p  z\, \left( x \m y \right) \, e^{z}
                \right] \; H   \nn \\
 && +\;  \left[ \left( y \m z \right) \, e^{x} \p   \left( z \m x \right)
            \, e^{y} \p \left( x \m y \right) \, e^{z} \right] \;  
H^2 \nn
\ena }}  \labelpr{su3expo}
\ene
\section{The exponential map of GL(N)\labelpr{secsunexp}}
As we have seen in the cases of the groups SU(3) and SU(2,2) \cite{bzl2} 
the exponential
map  can be written as sum over the first (N-1) powers of the generator 
$H \, \in $ \liealg{su}(N). Where the coefficients are functions of the 
eigenvalues of $H$. In this section we generalize the results
we have gotten on the low dimensional examples. It seems that there is a 
relatively easy concept of generalization.
\\
The desired result is an expansion of the exponential map of the form
\bee
\mbox{g} \abgl e^{H} \abgl \sum_{n=0}^{\infty} \, \frac{H^{n}}{n!}
\abgl  \sum_{k=0}^{N-1} \;  A_{k} \; H^{k}
\ene
where the coefficients $A_{k} $ are rational functions of the  
eigenvalues
$ \left\{ \lambda_{i} \; ; \; i \, =\, 1,2,\ldots , N \right\} $ of 
the generator $H$.
\subsection{The secular equation}
The first step will be to take a look at the eigenvalues, some
auxiliary functions, and their interrelations.
\\
Let us consider the secular equation of the matrix $H$
\bee
0 \abgl \prod_{i=1}^{N} \left( \lambda \m \lambda_{i} \right) \abgl
 \m \left( \sum_{k=0}^{N} C_{k}\; \lambda^{N-k}\right)  \, ,
 \labelpr{suNseceq}
\ene
where the coefficients $ C_{k} $ are functions of the eigenvalues of $H$.
In section \ref{misc} some coefficients are listed in their  
explicit form.
For later convenience we introduce also the ``truncated'' version
$C_{(i)k} $ of the coefficients $C_{k} $, defined by
\bee
\prod_{j\neq i} \left( \lambda \m \lambda_{j} \right) \quad =: \quad
 \m \sum_{k=0}^{N-1} C_{(i)k} \; \lambda^{N-1-k} \labelpr{suNseceqtrun} 
\ene
Essentially the $C_{(i)k} $ contain all terms of $C_{k} $ without
$\lambda_{i}$. The connection between these coefficients can be seen 
easily via
\goodbreak
\bea
\prod_{i=1}^{N}\, \left( \lambda \m \lambda_{i} \right) \tgl
\left( \lambda \m \lambda_{1} \right) \;
\prod_{i=2}^{N}\, \left( \lambda \m \lambda_{i} \right) \nn \\
\tgl \left( \lambda \m \lambda_{1} \right) \;
\left( \lambda^{N-1} \m \sum_{k=1}^{N-1} C_{(1)k} \; \lambda^{N-1-k}
\right) \nn \\
\tgl \lambda^{N}
   \p \lambda_{1} \, C_{(1)N-1} \m \sum_{k=0}^{N-2} \, \left[
  C_{(1)k+1} \m \lambda_{1} \, C_{(1)k} \right] \lambda^{N-1-k} \nn \\
&\gefordert & \lambda^{N} \m \sum_{k=0}^{N-1} \, C_{k+1} \;  
\lambda^{N-1-k}
\, . \nn
\ena
Since the calculations above can be generalized to all eigenvalues
we get the relations
\bea
C_{k} \tgl C_{(i)k} \m \lambda_{i} \,  C_{(i)k-1} \qquad \mbox{for} 
\quad k \agl 1,2,\ldots , N-1  \labelpr{suNcoeffrel} \\
C_N \tgl -\, \lambda_{i} \, C_{(i)N-1}  \, . \nn
\ena
Let us define the multiplier $m $; i.e., the discriminant of the
secular equation
\bee
m\abdef \prod_{i < j} \left( \lambda_{i} \m \lambda_{j} \right)
\labelpr{mdef}
\ene
and the functions (see also \verw{suNmiexpl})
\bee
m_{i} \abdef m\, \left (\lambda_1 \, , \, \ldots \, ,\,  
\lambda_{i-1} \, ,\,
             \lambda_{i+1}\, ,\, \ldots \, ,\, \lambda_{N} \right)
\abgl
\prod_{j < k \atop j,k \neq i} \left( \lambda_{j} \m \lambda_{k}  
\right)\, .
\ene
In what follows we mainly use the following form of $m$ which can be 
obtained by expanding the Slater determinant
(see \verw{slaterdet} and \verw{suNmexplpr})
\bee
m \abgl  \sum_{i=1}^{N} \; (-1)^{i+1} \; m_{i}\; {\lambda_{i}}^{N-1} 
    \, .    \labelpr{suNmexpl}
\ene
This formula can be generalized to
\bee
m\; \delta_{kl} \abgl  \sum_{i=1}^{N} \; (-1)^{i} \;  m_{i}
   \; C_{(i)k-1} {\lambda_{i}}^{N-l}  \qquad \mbox{for}
\quad k,l\, =\, 1,2,\ldots , N \; . \labelpr{suNmCrel}
\ene
For the proof see section \ref{secslater}.
\subsection{Recurrence relations}
The described method relies on the Cayley-Hamilton theorem which  
gives us the
ability to write all powers $H^{N+n} $ for $n \, \in $ \koer{N} in terms
of the first N-1 powers of $H$.
The Cayley-Hamilton theorem for $H\, \in $ \liealg{gl}(N) reads
\bee
H^{N} \abgl  \sum_{k=1}^{N} C_{k}\; H^{N-k} \labelpr{suNcayham}
\ene
The coefficients $C_{k}$ are the same as those in the secular equation
\verw{suNseceq} and satisfy the recurrence relations \verw{suNrecgen} 
derived below.
For the special groups, i.e., det g = 1 for g $\in $ SL(N) the first
coefficient vanishes since the sum over the eigenvalues is zero.
\\[10pt]
Multiplication of \verw{suNcayham} with $H^{n} $ and using  
\verw{suNcayham}
again gives the iterated form
\bee
H^{N+n} \abgl \sum_{k=1}^{N} C_{k}^{n} \; H^{N-k}
\ene
Multiplying once more with $H$ gives
\bea
H^{N+n+1} \tgl \left( C^{n}_{2} \p C^{n}_{1} \, C_{1} \right) H^{N-1} \p 
\left( C^{n}_{3} \p C^{n}_{1} \, C_{2} \right) H^{N-2} \nn \\
& + & \ldots \p
\left( C^{n}_{n+1} \p C^{n}_{1} \, C_{n} \right) H^{N-n}
\p \ldots \p \nn \\
& + &  \left( C^{n}_{N} \p C^{n}_{1} \, C_{N-1} \right) H
\p C^{n}_{1}\, C_{N}  \nn \\
&\gefordert & \sum_{k=1}^{N} C_{k}^{n+1} \; H^{N-k} \nn
\ena
and hence we get the recurrence relations
\bea
C^{n+1}_{1} \agl C^{n}_{2} \p C^{n}_{1} \, C_{1} \qquad && \qquad
C^{n+1}_{2} \agl C^{n}_{3} \p C^{n}_{1} \, C_{2} \nn \\
\ldots \qquad
C^{n+1}_{k} \tgl C^{n}_{k+1} \p C^{n}_{1} \, C_{k} \qquad \ldots
 \labelpr{suNrecgen} \\
C^{n+1}_{N-1} \agl C^{n}_{N} \p C^{n}_{1} \, C_{N-1} \qquad && \qquad
C^{n+1}_{N} \agl C^{n}_{1} \, C_{N} \nn \, .
\ena
If we successively plug in the $ C^{j}_{k} $ in the recurrence relation
of $C^{n}_{1} $
we get a formula which contains only terms with $ C_{1}^{n} $ and the 
coefficients of the original Cayley-Hamilton equation \verw{suNcayham}
\bee
C^{n+1}_{1} \agl  \left\{ \begin{array}{r@{\quad \mbox{for} \quad}l}
{\displaystyle \sum_{j=1}^{N}\; C^{n+1-j}_{1} \; C_{j}} &  n\, \geq  
\, N-1 \\
{\displaystyle \sum_{j=0}^{n}\; C_{1}^{n-j} \; C_{1+j}} \p m\; C_{n+2}  &
   n\, < \, N-1
\end{array} \right.   \labelpr{suNrec}
\ene
For the other coefficients $ C_{k}^{n+1} \; ( k \, =\, 1,2, \ldots  
, N ) $
we get analogous formulae
\bee
C^{n+1}_{k} \agl  \left\{ \begin{array}{r@{\quad \mbox{for} \quad}l}
{\displaystyle \sum_{j=0}^{N-k}\; C^{n-j}_{1} \; C_{k+j}} &
  n\, \geq \, N-k \\
{\displaystyle \sum_{j=0}^{n}\; C_{1}^{n-j} \; C_{k+j}} \p m\;  
C_{k+n+1}  &
   n\, < \, N-k
\end{array} \right.   \labelpr{suNreck}
\ene
The coefficients of the secular equation have the explicit form
\bea
m\; C_{k} \teqverw{suNmexpl}  \sum_{i=1}^{N} \; (-1)^{i+1}\; m_{i} \; 
   \lambda_{i}^{\, N-1} \; C_{k}  \eqverw{suNcoeffrel}
\sum_{i=1}^{N} \; (-1)^{i+1}\; m_{i} \; \lambda_{i}^{\, N-1}
   \left( C_{(i)k} \m \lambda_{i} \; C_{(i)k-1} \right) \nn \\
\tgl - \;  \sum_{i=1}^{N} \; (-1)^{i+1}\; m_{i} \; \lambda_{i}^{\, N}
     \; C_{(i)k-1}   \p \underbrace{\sum_{i=1}^{N} \; (-1)^{i+1}\; m_{i} 
     \; \lambda_{i}^{\, N-1}\;  C_{(i)k} }_{=\; 0 \quad \mbox{for}\; 
      k\, \neq \, 0 }  \nn
\ena
where we applied Eq.\verw{suNmCrel2} to the second term in the
last equation.
We will need this form as first values in the proof of  
Eq.\verw{suNCknexp}
\bee
\fbox{$ \displaystyle  \quad
 m\; C_{k} \abgl   \sum_{i=1}^{N} \; (-1)^{i}\; m_{i} \;
   \lambda_{i}^{\, N} \; C_{(i)k-1}   \quad
\mbox{for} \; k \, =\, 1,2, \ldots , N \quad $}  \labelpr{suNCkexp}
\ene
{}From the SU(3) and SU(2,2) cases one may assume that the recurrence 
relation \verw{suNrec} has the solution
\bee
m\; C^{n}_{1} \abgl \sum_{i=1}^{N} \; (-1)^{i+1} \; m_{i}\,
{ \lambda_{i}}^{N+n}  \, .  \labelpr{suNrec1}
\ene
{\bf Proof}:
\\[5pt]
The proof of Eq.\verw{suNrec1} is done by induction over $n$.
\\[5pt]
The first coefficient ($n$ = 0) is given by
\bee
m\; C_{1} \abgl \sum_{i=1}^{N} \; (-1)^{i+1} \; m_{i}\,
{ \lambda_{i}}^{N}
\ene
which is easy to prove if one writes $C_{1} $ in the form
\bdis
C_{1} \abgl \sum_{j=1}^{N}\; \lambda_{j} \agl \lambda_{i} \p
  \sum_{j\neq i} \; \lambda_{j} \agl  \lambda_{i} \p C_{(i)1} \, .
\edis
For the product $m\; C_{1} $ we take $m$ in the form of
Eq.\verw{suNmexplpr} we get
\bea
m\; C_{1} \tgl \sum_{i=1}^{N}\; (-1)^{i+1} \; m_{i}\,  
\lambda_{i}^{\; N-1}
\left(  \lambda_{i} \p C_{(i)1}  \right)  \nn \\
\tgl \sum_{i=1}^{N}\; (-1)^{i+1} \; m_{i}\, \lambda_{i}^{\; N} \p
\underbrace{\sum_{i=1}^{N}\; (-1)^{i+1} \; m_{i}\; C_{(i)1}\;
 \lambda_{i}^{\; N-1}}_{=\, 0}  \nn
\ena
The last equation holds since the exponent of $\lambda_{i} $ should be 
$N-2$ in order to yield a non-vanishing sum (see Eq.\verw{suNmCrel2}). 
\\[10pt]
First we treat the case of $n\, \geq \, N $.
The next step is to assume the validity of \verw{suNrec1} for $n$  
and to show
that then it follows also for $n+1$
\goodbreak
\bea
m\, C_{1}^{n+1} \teqverw{suNrec} \sum_{j=1}^{N} \; m \,  
C_{1}^{n+1-j}\, C_{j}
\eqverw{suNrec1} \sum_{j=1}^{N} \;\sum_{i=1}^{N} \; (-1)^{i+1} \;  
m_{i}\,
  { \lambda_{i}}^{N+n+1-j}  \, C_{j}   \nn \\
\teqverw{suNcoeffrel} \sum_{j=1}^{N} \;\sum_{i=1}^{N} \; (-1)^{i+1} 
   \; m_{i}\, { \lambda_{i}}^{N+n+1-j}
    \, \left( C_{(i)j} \m \lambda_{i} \, C_{(i)j-1} \right) \nn \\
\tgl \sum_{i=1}^{N}  \; (-1)^{i+1} \;  m_{i}\,
    \left( \sum_{j=1}^{N} \; C_{(i)j} \,
    { \lambda_{i}}^{N+n+1-j} \m  \sum_{j=1}^{N} \; C_{(i)j-1} \,
    { \lambda_{i}}^{N+n+2-j} \right)  \nn \\
\tgl \sum_{i=1}^{N} \; (-1)^{i+1} \; m_{i} \,
     \left( \sum_{j=2}^{N+1} \; C_{(i)j-1} \,
    { \lambda_{i}}^{N+n+2-j}  \m \sum_{j=1}^{N} \; C_{(i)j-1} \,
    { \lambda_{i}}^{N+n+2-j} \right)  \nn \\
\tgl \sum_{i=1}^{N}  \; (-1)^{i+1} \; m_{i}\,
     \left( \, \underbrace{C_{(i)N}}_{=\, 0} \,
   { \lambda_{i}}^{n+1} \m  \underbrace{C_{(i)0}}_{= \, -1 } \,
    { \lambda_{i}}^{N+n+1} \right)  \nn \\
\tgl  \sum_{i=1}^{N}  \; (-1)^{i+1} \; m_{i}\, { \lambda_{i}}^{N+n+1}  
 \abgl m\, C_{1}^{n+1}  \, . \nn
\ena
In the case of $ n\, < \, N-1 $ there is an additional term
\bea
m\, C_{1}^{n+1} \tgl \sum_{j=0}^{n} \; m \, C_{1}^{n-j}\, C_{1+j}
      \p m\, C_{n+2} \agl \ldots \agl \nn \\
\tgl \sum_{i=1}^{N}  \; (-1)^{i+1} \; m_{i}\, \left( C_{(i)n+1} \,
   \lambda_{i}^{\, N} \m C_{(i)0} \, \lambda_{i}^{\, N+n+1} \right)
    \p m\, C_{n+2}  \nn \\
\tgl  \sum_{i=1}^{N}  \; (-1)^{i+1} \; m_{i}\, \lambda_{i}^{\, N+n+1}
  \p  \underbrace{\sum_{i=1}^{N}  \; (-1)^{i+1} \; m_{i}\, C_{(i)n+1} 
   \, \lambda_{i}^{\, N}}_{= \, -\; m\, C_{n+2}}  \p m\, C_{n+2}  \nn \\
\teqverw{suNCkexp}
\sum_{i=1}^{N}  \; (-1)^{i+1} \; m_{i}\, \lambda_{i}^{\, N+n+1}  \, .\nn 
\ena
\qed
\\[10pt]
The coefficients $C_{k}^{n}$ can be written in the form
\bee
\fbox{$ \displaystyle  \quad
m\; C_{k}^{n} \abgl \sum_{i=1}^{N} \; (-1)^{i} \;  m_{i}
   \; C_{(i)k-1} {\lambda_{i}}^{N+n}  \qquad \mbox{for}
\quad k\, =\, 1,\ldots , N  \quad $}  \labelpr{suNCknexp}
\ene
{\bf Proof}
\\[5pt]
The proof is analogous to the one of $m\; C_{1}^{n} $ but uses the  
explicit
form \verw{suNrec1} of these coefficients.
\bea
m\; C_{k}^{n+1} \tgl \sum_{j=0}^{N-k} \; m\; C_{1}^{\; n-j} \; C_{k+j} 
\eqverw{suNrec1} \sum_{j=0}^{N-k} \; \sum_{i=1}^{N} \; (-1)^{i+1}\;
m_{i} \; \lambda_{i}^{N+n-j}\;   C_{k+j} \nn \\
\teqverw{suNcoeffrel} \sum_{j=0}^{N-k} \; \sum_{i=1}^{N} \; (-1)^{i+1}\;
  m_{i} \; \left( C_{(i)k+j} \m \lambda_{i} \;  C_{(i)k+j-1} \right) \, 
  \lambda_{i}^{N+n-j} \nn \\
\tgl \sum_{j=0}^{N-k} \; \sum_{i=1}^{N} \; (-1)^{i+1}\;
  m_{i} \;  C_{(i)k+j} \; \lambda_{i}^{N+n-j} \nn \\
&& \m  \sum_{j=1}^{N-k} \; \sum_{i=1}^{N} \; (-1)^{i+1}\;
  m_{i} \; C_{(i)k+j-1} \;  \lambda_{i}^{N+1+n-j} \nn \\
&& \m  \sum_{i=1}^{N} \; (-1)^{i+1}\;
  m_{i} \; C_{(i)k-1} \;  \lambda_{i}^{N+n+1}  \, . \nn
\ena
If we now shift the summation index $j$ in the second sum most of  
the terms
cancel with those of the first sum. The remaining term $j\, =\, N-k $ 
in the first sum contains $ C_{(i)N} \agl 0 $, hence, vanishes also. 
Therefore, only the third sum remains what proofs the
assumed form \verw{suNCknexp} of $C_{k}^{n}$.
\\
Again there are the cases $n\, <\, N-k $ which need to be treated  
separately
\bea
m\; C_{k}^{n+1} \tgl \sum_{j=0}^{N-k} \; m\; C_{1}^{\; n-j} \; C_{k+j} 
  \p m\; C_{k+n+1}   \agl \ldots \agl        \nn \\
\tgl  \sum_{i=1}^{N} \; (-1)^{i}\; m_{i} \;  C_{(i)k-1} \;
  \lambda_{i}^{N+n+1} \p \underbrace{\sum_{i=1}^{N} \; (-1)^{i+1}\; m_{i}
  \;  C_{(i)k+n} \; \lambda_{i}^{N}}_{=\; -\, m\; C_{k+n+1}
   \quad \verw{suNCkexp} } \p
  m\; C_{k+n+1} \nn \\
\tgl  \sum_{i=1}^{N} \; (-1)^{i}\; m_{i} \;  C_{(i)k-1} \;
  \lambda_{i}^{N+n+1} \nn  \, .
\ena
\qed
\subsection{The exponential map}
The expansion of a group element g $ \in $ G with generator
$ H \, \in $ \liealg{gl}(N) can now be written like
\bea
\mbox{g} \agl e^H \tgl \sum_{n=0}^{\infty} \, \frac{H^{n}}{n!}
\abgl \sum_{n=0}^{N-1} \, \frac{H^{n}}{n!}  \p \sum_{n=0}^{\infty} \,
   \frac{H^{N+n}}{(N+n)!}  \nn \\
\tgl \sum_{n=0}^{N-1} \, \frac{H^{n}}{n!}  \p \sum_{n=0}^{\infty} \,
   \frac{1}{(N+n)!}\, \left( \sum_{k=1}^{N}\; C_{k}^{n}  \;  
H^{N-k}\right)
 \; . \labelpr{suNexp1}
\ena
Using the multiplier $m$ we get
\bee
m\, \mbox{g} \agl m\;   \sum_{n=0}^{N-1} \, \frac{H^{n}}{n!} \p
   \sum_{k=1}^{N}\; \left[ \sum_{n=0}^{\infty} \,  \frac{1}{(N+n)!}\, 
    m\, C_{k}^{n} \,  \right] \; H^{N-k}  \, .\labelpr{suNexp2}
\ene
We can now treat the sums for different $k$ separately
\bea
\sum_{n=0}^{\infty} \,  \frac{1}{(N+n)!}\,  m\, C_{k}^{n}  \tgl
\sum_{n=0}^{\infty} \,  \frac{1}{(N+n)!}\,  \sum_{i=1}^{N} \;
        (-1)^{i} \; m_{i}\; C_{(i)k-1} \; \lambda_{i}^{N+n} \nn \\
\tgl  \sum_{i=1}^{N} \;  (-1)^{i} \; C_{(i)k-1} \; m_{i}\;
        \sum_{n=0}^{\infty} \; \frac{1}{(N+n)!}\,    
\lambda_{i}^{N+n}  \nn \\
\tgl \sum_{i=1}^{N} \; (-1)^{i} \;  C_{(i)k-1} \;  m_{i}\;
  \left(  e^{\lambda_{i}} \m \sum_{n=0}^{N-1}
  \, \frac{\lambda_{i}^{\; n}}{n!} \right) \nn \\
\tgl  \sum_{i=1}^{N} \; (-1)^{i} \; C_{(i)k-1} \; m_{i}\;
    e^{\lambda_{i}}  \p
   \sum_{n=0}^{N-1}\; \frac{1}{n!} \; \sum_{i=1}^{N} \; (-1)^{i+1} \; 
     C_{(i)k-1} \; m_{i}\; \lambda_{i}^{\; n} \nn \\
\tgl \sum_{i=1}^{N} \; (-1)^{i} \; C_{(i)k-1} \;  m_{i}\;   
e^{\lambda_{i}}
   \m   \frac{m}{(N-k)!}   \nn
\ena
The last equation relies on Eq.\verw{suNmCrel1} and  \verw{suNmCrel2}.
The terms $ \m \frac{m}{(N-k)!} $ cancel the first sum in  
Eq.\verw{suNexp2}.
\\[10pt]
The final result turns out to be
\bea
m\, e^H \tgl (-1)^{N} \; \det H \, \left( \sum _{i=1}^{N} \; (-1)^{i} \; 
   m_{i} \; \frac{e^{\lambda_{i}}}{\lambda_i}\right)  \, I_N  \p
 \left( \sum _{i=1}^{N} \;  (-1)^{i} \;  C_{(i)N-2}\; m_{i}
     \; e^{\lambda_i}  \right) \, H \nn \\
&& + \; \ldots \p  \left( \; \sum _{i=1}^{N} \;  (-1)^{i} \; C_{(i)k}\; 
 m_{i} \; e^{\lambda_i} \right) \, H^{N-1-k} \p \ldots \nn \\
&& +\;  \left( \; \sum _{i=1}^{N} \;(-1)^{i} \;
\; m_{i} \; \lambda_{i}\;  e^{\lambda_{i}} \, \right) \, H^{N-2}
\p  \left( \sum _{i=1}^{N} \; (-1)^{i+1} \; m_{i}
\; e^{\lambda_{i}} \, \right)  \, H^{N-1} \nn
\ena
or in closed form
\bee
\fbox{$ \displaystyle
m\, e^{H} \abgl \sum_{n=1}^{N} \; \left( \sum_{i=1}^{N} \; (-1)^{i} \;
 C_{(i)n-1} \; m_{i} \;  e^{\lambda_{i}}  \right) \; H^{N-n}
 $ }    \labelpr{suNfinal}
\ene
which reads in terms of the adjoints of  the Slater determinant
\bee
\fbox{$ \displaystyle
m\, e^{H} \abgl \sum_{n=0}^{N-1} \; \left( \sum_{i=1}^{N} \;  A_{(i)n} 
  \; e^{\lambda_{i}}  \right) \; H^{n} $ } \labelpr{suNfinal1}
\ene
\subsection{The Slater determinant \labelpr{secslater}}
One crucial ingredient of the method is the usage of a multiplier
$m$, defined in Eq.\verw{mdef}. {}From low dimensional examples one 
may assume the forms \verw{suNmexpl} and \verw{suNmCrel} of $m$.
The general proofs can be done by writing $m$ as Slater determinant. 
The Slater determinant \index{Slater determinat} is defined as
(cf. \cite{fubini, itzykdrouffe})
\bee
\left| \matrix{ 1 & 1 & \ldots & 1 \cr
\lambda_N & \lambda_{N-1} & \ldots & \lambda_{1} \cr
\lambda_N^{2}  & \lambda_{N-1}^{2}  & \ldots & \lambda_{1}^{2}  \cr
\vdots & \vdots & \ddots & \vdots \cr
\lambda_N^{N-1}  & \lambda_{N-1}^{N-1}  & \ldots & \lambda_{1}^{N-1} \cr
} \right|
\abgl \prod_{i<j} \; \left( \lambda_{i} \m
 \lambda_{j} \right) \abgl m \, .\labelpr{slaterdet}
\ene
We can now use the Laplacian method of expanding the Slater determinant
\bee
\det A \abgl \sum_{j=1}^{n} a_{ij} \; A_{ij}
       \abgl \sum_{i=1}^{n} a_{ij} \; A_{ij}   \labelpr{laplace}
\ene
where the so-called adjoints $  A_{ij} $ are the subdeterminants
of $ a_{ij} $ multiplied by the sign factor $(-1)^{i+j}$.
\\
It is also well known that the
Laplace expansion with ``wrong'' adjoints gives zero
\bee
0 \abgl \sum_{j=1}^{n} a_{ij} \; A_{lj}  \qquad \mbox{for}
  \quad l \, \neq \, i  \, .   \labelpr{laplacewrong}
\ene
The Laplacian method applied  with respect to the last
row gives then the expansion \verw{suNmexpl}
\bea
m \tgl \sum_{i=1}^{N} \; \lambda_{i}^{N-1} \; A_{(N+1-i)N}
  \agl \sum_{i=1}^{N} \; (-1)^{N + (N+1-i)} \; m_{i} \;  
\lambda_{i}^{N-1}
\nn \\
\tgl \sum_{i=1}^{N} \; (-1)^{i+1} \; m_{i} \; \lambda_{i}^{N-1}
\, , \labelpr{suNmexplpr}
\ena
where $m_{i} $ are the subdeterminants of $m $
\bee
m_{i} \abgl \prod_{k<j \atop k,j \neq i }   \; \left( \lambda_{k} \m
 \lambda_{j} \right) \, . \labelpr{suNmiexpl}
\ene
We get
\bea
m \tgl m_{i} \;  \prod_{k<i}\; \left( \lambda_{k} \m \lambda_{i} \right) 
 \; \prod_{i<j}\; \left( \lambda_{i} \m \lambda_{j} \right)
   \qquad \mbox{for} \quad i\, =\, 1,2,\ldots ,N \nn \\
\tgl (-1)^{i-1} \; m_{i} \; \prod_{j\neq i} \;
                    \left( \lambda_{i} \m \lambda_{j} \right)  \nn \\
\teqverw{suNseceqtrun} (-1)^{i-1} \; m_{i}\; \sum_{n=0}^{N-1} \;
  \left( -\; C_{(i)n} \; \lambda_{i}^{\, N-1-n}  \right) \nn \\
\tgl \sum_{n=0}^{N-1} \; (-1)^{i} \; m_{i}\;  C_{(i)N-1-n} \;
  \lambda_{i}^{\, n}
\eqverw{laplace} \sum_{n=0}^{N-1} \; A_{(i)n} \; \lambda_{i}^{\, n}
\ena
what proves Eq.\verw{suNmCrel1}. Also Eq.\verw{suNmCrel2} is proven since
if the exponent of $\lambda_{i} $ is not $n$ the sum vanishes  
because it is
an expansion with the ``wrong'' adjoints $A_{(i)n}$.
%
%
Hence we get an explicit expression for the adjoints
\bee
A_{(i)n} \abgl (-1)^{i} \; m_{i} \; C_{(i)N-n-1} \, . \labelpr{suNadj}
\ene
We can also expand $m$ with respect to the (n+1)-th line and then use
Eq.\verw{suNadj}
\bea
m \tgl  \sum_{i=1}^{N} \; A_{(i)n} \; \lambda_{i}^{\, n}
        \qquad \mbox{for} \quad n\, =\, 0,1,\ldots ,N-1  \nn \\
\tgl \sum_{i=1}^{N} \;  (-1)^{i} \; m_{i}\;  C_{(i)N-1-n} \;
       \lambda_{i}^{\, n}  \, . \labelpr{suNmCrel1}
\ena
Writing the Laplacian expansion with ``wrong'' adjoints leads to
\bea
0 \tgl \sum_{i=1}^{N} \; A_{(i)k} \; \lambda_{i}^{\, n}
        \qquad \mbox{for} \quad k,n\, =\, 0,1,\ldots ,N-1
        \quad k\, \neq \, n  \nn \\
\tgl \sum_{i=1}^{N} \;  (-1)^{i} \; m_{i}\;  C_{(i)N-1-k} \;
       \lambda_{i}^{\, n}  \labelpr{suNmCrel2}
\ena
%
Ergo (cf. Eq.\verw{suNmCrel})
\bee
\fbox{$ \displaystyle
m\; \delta_{kl} \abgl  \sum_{i=1}^{N} \; (-1)^{i} \;  m_{i}
   \; C_{(i)k-1} {\lambda_{i}}^{N-l}  \qquad \mbox{for}
\quad k,l\, =\, 1,2,\ldots , N   $}
\ene
\\[20pt]
\begin{appendix}
{\Large\bf Appendix}\labelpr{misc}
\section{Some details}
This section contains explicit forms of some coefficients and some  
proofs.
Almost all of the equations hold
in the general case, but those which hold only in the case of the  
special
groups, i.e., vanishing sum of eigenvalues, are denoted by the sign  
$\Sequ $.
\\[10pt]
For the coefficients of the secular equation we get, e.g.,
\bea
C_{N} \tgl (-1)^{N+1} \; \prod_{i=1}^{N} \, \lambda_{i} \abgl
   (-1)^{N+1} \; \det H  \nn \\
C_{N-1} \tgl (-1)^{N} \; \sum_{i=1}^{N} \;
        \prod_{j \neq i}\;  \lambda_{j} \abgl    (-1)^{N} \;
\sum_{i=1}^{N} \; \frac{\det H}{\lambda_{i}} \labelpr{suNchcoeff} \\
C_{2} \tgl (-1) \; \sum_{i<j} \, \lambda_{i}  \, \lambda_{j} \Sequ
   \frac{1}{2} \; \sum_{k=1}^{N} \; {\lambda_{k}}^{2} \nn \\
C_{1} \tgl  \sum_{i=1}^{N} \; \lambda_{i} \Sequ 0 \; , \qquad \qquad
C_{0} \agl -\, 1  \,  .\nn
\ena
Some ``truncated'' coefficients
\bea
C_{(i)N-1} \tgl (-1)^{N-2} \; \prod_{j\neq i} \, \lambda_{j} \agl
   (-1)^{N} \; \frac{\det H }{\lambda_{i}} \nn \\
C_{(i)N-2} \tgl (-1)^{N-3} \; \sum_{k\neq i} \;
                  \prod_{j \neq k, i}\;  \lambda_{j}  \nn \\
C_{(i)2} \tgl (-1) \; \sum_{j<k \atop j,k\neq i}
 \, \lambda_{j}  \, \lambda_{k} \; , \qquad   \qquad
C_{(i)1} \agl  \sum_{j\neq i} \, \lambda_{j} \Sequ - \, \lambda_{i}  
\nn \\
C_{(i)0} \tgl  -\, 1   \; , \qquad \qquad
C_{(i)N} \agl 0   \, . \nn
\ena
\section{Additional checks}
\subsection{One-dimensional subgroups}
One-dimensional subgroups (cf. \cite{helga}) of GL(N) can be generated by
\bee
\left\{ e^{t\, H} \; ; \; H\, \in \, \mbox{\liealg{gl}(N)} \; , \;
        t\, \in\, \koer{R} \right\} \, .
\ene
{}From the known expansion of $e^{H} $ we can derive the expansion of 
$e^{t\, H}$ by multiplying the occurring expressions with an appropriate 
factor. Obviously, the eigenvalues of the t-dependent generator  
$t\, H $ are
$t\, \lambda_{i} $ if the $\lambda_{i} $ are the eigenvalues of $H$.
Therefore, we need to make the replacements
\bea
\lambda_{i} & \longrightarrow & t\, \lambda_{i}   \nn \\
C_{k}       & \longrightarrow & t^{k} \, C_{k} \hskip 60pt
C_{(i)k}    \; \longrightarrow \; t^{k} \, C_{(i)k}   \nn \\
m           & \longrightarrow & t^{N(N-1)/2} \, m  \hskip 30pt
m_{i} \quad \longrightarrow \; t^{(N-1)(N-2)/2} \, m_{i}  \nn
\ena
The expansion \verw{suNfinal1} reads now
\bea
 t^{N(N-1)/2} \; m \; e^{t\, H} \tgl   \sum_{n=0}^{N-1}\;
   \left( \sum_{i=1}^{N} \; (-1)^{i} \; t^{n-1} \, C_{(i)n-1} \;
    t^{(N-1)(N-2)/2}  \, m_{i}  \; e^{t\, H} \right)
       \left( t\, H \right) ^{N-n}  \nn \\
\tgl t^{N(N-1)/2} \, \sum_{n=0}^{N-1}\;
   \left( \sum_{i=1}^{N} \; (-1)^{i} \; C_{(i)n-1} \;
   \, m_{i}  \; e^{t\, H} \right) \;  H ^{N-n}  \nn
\ena
or
\bee
m \; e^{t\, H} \abgl   \sum_{n=0}^{N-1}\; \left( \sum_{i=1}^{N} \;
  (-1)^{i} \; C_{(i)n-1} \; \, m_{i}  \; e^{t\, H} \right) \;   
H^{N-n} \, .
\labelpr{suNtHexp}
\ene
Differentiation of the r.h.s of Eq.\verw{suNtHexp} and setting $  
t\, =\, 0$
gives the derivation of the unit element
\bdis
\sum_{n=0}^{N-1}\; \Bigl( \;  \underbrace{ \sum_{i=1}^{N} \;
  (-1)^{i} \; C_{(i)n-1} \; \, m_{i}  \; \lambda_{i} }_{m\,  
\delta_{n,N-1}
  \;  (cf. \verw{suNmCrel}) }  \; \Bigr) \;  H^{N-n}  \abgl m\; H  \, .
\edis
Since this result coincides with the one we get by differentiating the 
l.h.s  it is an additional proof of the expansion \verw{suNfinal}.
\subsection{Eigenvalues}
It is easy to demonstrate that Eq.\verw{suNfinal1} gives also the right
connection between the eigenvalues of the generator $H$ and the  
ones of the
corresponding group element $g\, =\, e^{H}$. Let $x_{i} $ be the  
eigenvectors
of $H$ with eigenvalues $\lambda_{i} $
\bdis
H\; x_{i} \abgl \lambda_{i}\; x_{i} \qquad \mbox{for} \quad
      i\, =\, 1,2, \ldots ,N \, .
\edis
For the powers of $H$ we get
\bdis
H^{n} \; x_{i} \abgl \lambda_{i}^{\, n} \; x_{i} \qquad \mbox{for} \quad 
  n\, \in \, \koer{N}  \, .
\edis
Plugging this in Eq.\verw{suNfinal} yields
\bea
m\; e^{H} \; x_{j} \tgl \sum_{n=0}^{N-1} \; \left( \sum_{i=1}^{N} \; 
 A_{(i)n} \;  e^{\lambda_{i}} \right) \; H^{n} \; x_{j}
\abgl \sum_{n=0}^{N-1} \; \left( \; A_{(i)n} \; e^{\lambda_{i}} \;
  \lambda_{i}^{\, n} \right) \; x_{j}  \nn \\
\tgl \sum_{i=1}^{N} \; \Bigl( \; \underbrace{\sum_{n=0}^{N-1} \;
  A_{(i)n} \;  \lambda_{i}^{\, n} }_{\quad m\; \delta_{ij} \;
 (cf. \verw{suNmCrel})} \; \Bigr) \; x_{j}
\abgl m\; e^{\lambda_{j}} \; x_{j} \, . \nn
\ena
Therefore, we get the desired result
\bdis
g\; x_{j} \abgl e^{\lambda_{j}} \; x_{j}
\edis
which again confirms the expansion \verw{suNfinal}.
\subsection{Remark \labelpr{suNrem1}}
In the cases where some of the eigenvalues coincide, the multiplier $m$
will be zero. But in these cases $m$ can be chosen in a simpler  
fashion so
that there occur only non-vanishing factors. Essentially all  
factors which
will become zero can be canceled out in Eq.\verw{suNfinal}.
\\[20pt]
\end{appendix}
{\Large \bf Acknowledgement}
\\[5pt]
The author would like to thank Dr. K-P. Marzlin for the
discussions of some points.

\begin{thebibliography}{10}

\bibitem{bzl1}
A.O. Barut, J.R. Zeni, and A.~Laufer.
\newblock {\em The exponential map for the conformal group {O(2,4)}}.
\newblock J.Phys.A:Math.Gen., 27:5239--5250, 1994.

\bibitem{bzl2}
A.O. Barut, J.R. Zeni, and A.~Laufer.
\newblock {\em The exponential map for the unitary group {SU(2,2)}}.
\newblock J.Phys.A:Math.Gen., 27:6799--6805, 1994.

\bibitem{chengli}
Ta-Pei Cheng and Ling-Fong Li.
\newblock {\em Gauge theory of elementary particle physics}.
\newblock Oxford University Press, 1984.

\bibitem{fubini}
Sergio Fubini.
\newblock {\em Vertex operators and {Q}uantum {H}all {E}ffect}.
\newblock Mod. Phys. Lett., A6: 347, 1991.

\bibitem{helga}
Sigurdur Helgason.
\newblock {\em Differential Geometry, Lie Groups, and Symmetric Spaces}.
\newblock Academic Press, 1978.

\bibitem{itzykdrouffe}
Claude Itzykson and Jean-Michel Drouffe.
\newblock {\em Statistical Field Theory II}.
\newblock Cambridge University Press, 1989.

\bibitem{zeni90}
J.Ricardo Zeni and Waldyr~A. Rodrigues.
\newblock Hadronic Journal, 13:317, 1990.

\bibitem{zeni92}
J.Ricardo Zeni and Waldyr~A. Rodrigues.
\newblock {\em A thoughtful study of {L}orentz transformations by  
{C}lifford
  algebras}.
\newblock International Journal of Modern Physics A, 7(8):1793--1817,
  1992.

\end{thebibliography}
\end{document}